\documentclass{jnmp}
\usepackage{amsmath}
\renewcommand{\d}{\operatorname{d}}

\JNMPnumberwithin{equation}{section}
\newtheorem{theorem}{Theorem}
\newtheorem{lemma}{Lemma}
\theoremstyle{definition}

\newtheorem*{remark}{Remark}
\newtheorem{example}{Example}
\begin{document}

\renewcommand{\evenhead}{L~Mart\'{\i}nez Alonso and  A~B~Shabat}
\renewcommand{\oddhead}{Towards a  Theory of Differential Constraints of a
Hydrodynamic  Hierarchy}

\thispagestyle{empty}

\FirstPageHead{10}{2}{2003}{\pageref{martinez-firstpage}--\pageref{martinez-lastpage}}{Article}

\copyrightnote{2003}{L~Mart\'{\i}nez Alonso and  A~B~Shabat}

\Name{Towards a  Theory of Differential Constraints\\
of a Hydrodynamic  Hierarchy}
\label{martinez-firstpage}

\Author{L~MART\'{I}NEZ ALONSO~$^\dag$ and A~B SHABAT~$^\ddag$}

\Address{$^\dag$ Departamento de F\'{\i}sica Te\'{o}rica II, Universidad
Complutense, E28040 Madrid, Spain \\
~~E-mail: luism@fis.ucm.es\\[10pt]
$^\ddag$~Landau Institute for Theoretical Physics, RAS,
Moscow 117 334, Russia  \\
~~E-mail: shabat@itp.ac.ru}

\Date{Received July 9, 2002; Revised October 2, 2002;
Accepted October 3, 2002}

\begin{abstract}
\noindent
We present a theory of compatible differential constraints of a
hydrodynamic hierarchy of infinite-dimensional systems. It
provides a convenient point of view for studying and formulating
integrability properties and it reveals some hidden structures of
the theory of integrable systems. Illustrative examples and new
integrable models are exhibited.
\end{abstract}

\section{Introduction}

In this paper we are concerned  with the hierarchy of evolution
equations
\begin{equation}\label{1i}
\frac{\partial G}{\partial t_i}=\langle A_i,G\rangle, \qquad
G=G(\lambda,x,{\boldsymbol{t}}),\quad {\boldsymbol{t}}:=(t_1,t_2,\ldots), \quad i\geq 1,
\end{equation}
where $\langle U,V\rangle:=UV_x-U_xV$, and the function $ G $ is
assumed to posses an expansion
\begin{equation}\label{2i}
G=g_0(x,{\boldsymbol{t}})+\frac{g_1(x,{\boldsymbol{t}})}{\lambda}+\frac{g_2(x,{\boldsymbol{t}})}{\lambda^2}+\cdots,\qquad
\lambda\rightarrow\infty,
\end{equation}
We consider two different forms of \eqref{1i} corresponding to
the cases $g_0\equiv 1$ and $g_0\not\equiv\mbox{const}$, which
will be henceforth referred to as the \emph{normalized} $G^{(n)}$
and the \emph{Schwarzian} $G^{(s)}$ hierarchies, respectively.
They are determined by two different definitions of the functions
$A_i=A_i(\lambda,x,{\boldsymbol{t}})$; namely,
\[
A_i^{(n)}=\lambda^i+\cdots+g_{i-1}\lambda+g_i,\qquad
A_i^{(s)}=g_0\lambda^i+\cdots+g_{i-1}\lambda,
\]
respectively. In both cases we may write
\[
A_i=(\lambda^i G)_+,\qquad i\geq 1,
\]
where the projectors $P:A\rightarrow A_+$ acting on power series
$A=\sum\limits_n a_n\lambda^n$ is defined as
\[
A_+^{(n)}:=P^{(n)}(A)=\sum_{n\geq 0} a_n\lambda^n, \qquad
A_+^{(s)}:=P^{(s)}(A)=\sum_{n\geq 1} a_n\lambda^n.
\]
In terms of the coefficients of the expansion of $G$, the
hierarchy \eqref{1i} becomes a set infinite-dimensional systems of
hydrodynamic type
\begin{gather*}
\partial_i g_n=\sum_{k=1}^n \langle g_{n-k}, g_{i+k}\rangle,\qquad
n\geq 0,\quad (\mbox{normalized case}),\\
\partial_i g_n=\sum_{k=0}^n \langle g_{n-k}, g_{i+k}\rangle,\qquad
n\geq 0,\quad (\mbox{Schwarzian case}).
\end{gather*}
We will show below that the case $g_0\not\equiv\mbox{const}$ can
be considered as the \emph{Schwarzian form} of the case $g_0\equiv
1$ and that a transformation connecting both cases exists.

Although equations similar to \eqref{1i} were considered in the
literature \cite{martinez:1,martinez:2,martinez:3,martinez:4} (starting with~\cite{martinez:1}), the
normalized case of the hierarchy \eqref{1i} was recently
reconsidered in~\cite{martinez:5} from the point view of the reduction
theory of integrable systems. It was in this context where the
notion of \emph{differential constraint} was introduced. One of
the main aims of the present work is to show that this notion is a
useful one in the theory of integrable systems.

For several purposes it is convenient to use instead of $G$ the
function
\begin{gather}
 H:=\frac{1}{G}=h_0+\frac{h_1}{\lambda}+\frac{h_2}{\lambda^2}+\cdots,\nonumber\\
 h_0=\frac{1}{g_0},\qquad h_1=-\frac{g_1}{g_0^2},\qquad
h_2=\frac{g_1^2}{g_0^3}-\frac{g_2}{g_0^2}, \qquad \ldots.\label{3i}
\end{gather}
It allows us to write \eqref{1i} in the alternative form
\begin{equation}\label{4i}
\frac{\partial H}{\partial t_i}=\frac{\partial }{\partial
x}\left(A_iH\right),
\end{equation}
which implies that  the coefficients of the expansion of $H$ in
powers of $\lambda$ supply an infinite set of conservation laws
for~\eqref{1i}.

 The function $H$ is particularly useful to
deal with the reductions of~\eqref{1i} determined by differential
constraints. Let us illustrate this fact by considering the first
flow $t:=t_1$ of the Schwarzian hierarchy, which according to
\eqref{4i} reads
\begin{equation}\label{5i}
h_{n,t}=\hat{h}_{n+1,x},\qquad \hat{h}_n:=\frac{h_n}{h_0},\qquad
n\geq 0.
\end{equation}
Let us look for a differential constraint of the form
\begin{equation}\label{6i}
\hat{h}_1=B(h_0,h_{0,x}),
\end{equation}
allowing to reduce the first equation in \eqref{5i} to
\begin{equation}\label{7i}
 h_{0,t}=D_x B(h_0,h_{0,x}),
\end{equation}
where $D_x$ stands for the total differentiation with respect
to~$x.$ The differential constraint~\eqref{6i} should be compatible
with the whole hierarchy. In particular, the second equation in~\eqref{5i} implies
\begin{equation}\label{8i}
D_t(h_1)=D_t(h_0B(h_0,h_{0,x})) \in \mbox{Im} \, D_x.
\end{equation}
Here $D_t$ is the total differentiation operator with respect to
$t$ and $\mbox{Im} D_x$ denotes the  range of $D_x$ acting on the
set of polynomials in the variables $h_0$, $h_{0,x}$, $h_{0,xx}$, $\ldots$.
Now one easily  finds that
\[
D_t(h_1)\equiv -(h_0B_{11}B_1)h_{0,xx}^2+\cdots,
\]
where $B_{11}$ is the second derivative of $B$ with respect to the
argument  $h_{0,x}$. Thus~\eqref{8i} yields $B_{11}=0$ and
therefore the constraint~\eqref{6i} should be of the form
\[
\hat{h}_1=b(h_0)+a(h_0)h_{0,x}.
\]
One can directly prove in this way that the only constraint
\eqref{6i} compatible with~\eqref{5i}) is
\begin{equation}\label{9i}
(\log h_0)_x=h_1.
\end{equation}
The corresponding reduced evolution equation~\eqref{7i} is given
by
\[
h_{0,t}+\left(\frac{1}{h_0}\right)_{xx}=0.
\]

  Several interesting generalizations of this type of constraints will be discussed in this paper.
For example it will be proved that the constraints
\begin{equation}\label{10i}
 (\log
h_0)_x=h_m,\qquad m\geq 1,
\end{equation}
and
\begin{equation}\label{11i}
 \{h_0,x\}=\left[H^2\right]_n,\qquad n\geq 1,
\end{equation}
are compatible with (\ref{5i}).  Here
\[
\{a,\, x\}:=
\frac34\left(\frac{a_x}{a}\right)^2-\frac12\frac{a_{xx}}{a}.
\]
and $\left[H^2\right]_n$ stands for the  coefficient at $\lambda^{-n}$ of the
expansion of $H^2$ in powers of $\lambda$.

The paper is organized as follows. In Section 2 we deal with the
structure of symmetries of the hierarchies \eqref{1i} and
reformulate some basic results. A particular attention is devoted
to applying the classical Liouville method of solution for solving
the polynomial reductions of \eqref{1i}. Some apparently new
results concerning $\tau$-functions like objects (Theo\-rem~2) and a
map $G^{(s)}\rightarrow G^{(n)}$ (Theorem 3) are also presented.
Section 3 is concerned with the use of the method of differential
constraints to characterize nonlinear integrable models from
\eqref{1i}. It is shown how the structure of wide classes of these
models as well as simple strategies for finding solutions can be
conveniently described by this method. Some new integrable models,
including a $2+1$-dimensional one \eqref{37} are exhibi\-ted.

\section{Symmetries}

 We first reformulate some already known basic properties about the symmetries
of the hierarchy~\eqref{1i}. Our starting point is the following
theorem~\cite{martinez:5}

\begin{theorem}
The evolutionary flows defined by \eqref{1i} form a commuting
family.
\end{theorem}

\begin{proof}
By using the Jacobi identity for the wronskian operation $\langle
U,V\rangle$ we have
\[
\frac{\partial^2 G}{\partial t_i\partial t_j}-\frac{\partial^2
G}{\partial t_j\partial t_i}=\left\langle\frac{\partial A_j}{\partial
t_i}-\frac{\partial A_i}{\partial t_j}+\langle
A_j,A_i\rangle,G\right\rangle.
\]

>From \eqref{1i} and taking into account that $A_i=(\lambda^i G)_+$
it follows
\begin{equation}\label{3}
\frac{\partial A_j}{\partial t_i}-\frac{\partial A_i}{\partial
t_j}=\left(\langle A_i,\lambda^j G\rangle  -\langle A_j,\lambda^i
G\rangle \right)_+.
\end{equation}
On the other hand
\[
\langle A_j,\lambda^i G\rangle=-\langle \lambda^i G,\lambda^j
G-(\lambda^j G)_- \rangle=\langle \lambda^i G,(\lambda^j
G)_-\rangle,
\]
where $(\lambda^j G)_-:=\lambda^j G-(\lambda^j G)_+$. Hence
\begin{gather*}
\frac{\partial A_j}{\partial t_i}-\frac{\partial A_i}{\partial
t_j}=\left(\langle A_i,\lambda^j G\rangle-\langle \lambda^i
G,(\lambda^j G)_-\rangle\right)_+ =\left(\langle A_i,\lambda^j
G\rangle-\langle A_i,(\lambda^j G)_-\rangle\right)_+\\
\phantom{\frac{\partial A_j}{\partial t_i}-\frac{\partial A_i}{\partial t_j}}{}
=\langle A_i,A_j\rangle_+=\langle A_i,A_j\rangle_,
\end{gather*}
which proves the statement.
\end{proof}

The hydrodynamic character of the equations of the
hierarchy~\eqref{1i} has important consequences. Thus, it follows that if
$\lambda=\lambda_0$ is either a zero or a pole of $G$ then
\[
\frac{\partial \lambda_0}{\partial t_i}=
A_i(\lambda_0,x,{\boldsymbol{t}})\frac{\partial \lambda_0}{\partial x}.
\]
It means that zeros and poles of $G$ are Riemann invariants
of~\eqref{1i}. This is a useful property because an obvious reduction
of~\eqref{1i} arises when only a finite number of coefficients in
the expansion~\eqref{2i} of $G$ are different from zero
\begin{equation}\label{6}
G^{(n)}=1+\frac{g_1}{\lambda}+\cdots+\frac{g_N}{\lambda^N},\qquad
G^{(s)}=g_0+\frac{g_1}{\lambda}+\cdots+\frac{g_{N-1}}{\lambda^{N-1}}.
\end{equation}
In this case it is possible to perform the integration of
\eqref{1i} in terms of Riemann invariants.  Let us illustrate this
feature for the normalized case. We can rewrite $G$ in the form
\begin{equation}\label{7}
G=\frac{1}{\lambda^N}\prod_{i=1}^N(\lambda+\gamma_i(x,{\boldsymbol{t}})),
\end{equation}
where $\gamma_i(x,{\boldsymbol{t}})$ are the zeros of $G$. Thus, under the
change of dependent variables
\[
\vec{g}:=(g_1,\dots,g_N)\mapsto
\vec{\gamma}:=(\gamma_1,\dots,\gamma_N),
\]
the system \eqref{1i} reduces to
\begin{equation}\label{8}
\partial_n\gamma_i=\Omega_{ni}(\vec{\gamma})\partial_x\gamma_i,\qquad
n\geq 1.
\end{equation}
Here $\Omega_{ni}:=A_n|_{\lambda=-\gamma_i}$ are given by
\begin{equation}\label{9}
\Omega_{ni}=\left\{\begin{array}{ll} (-1)^n\sum\limits_{j_l\neq
i}\gamma_{j_1}\gamma_{j_2}\cdots\gamma_{j_n}&
\mbox{if} \ \ n=1,2,\ldots,N-1,\vspace{1mm}\\
0&\mbox{if} \ \ n\geq N.
\end{array}
\right.
\end{equation}

The equations \eqref{8}--\eqref{9} form  a set of $N-1$ weakly
nonlinear hydrodynamic systems of Dubrovin type~\cite{martinez:2} which can
be integrated by means of a version of the classical Liouville
method~\cite{martinez:6} proposed by Tsarev~\cite{martinez:7} (see also~\cite{martinez:8}).
It basically consists in finding a vector field $D_x$ acting on
the new dependent variables $\vec{\gamma}$
\begin{equation}\label{10}
D_x\gamma_i=X_i(\vec{\gamma}),\qquad i=1,\ldots,N,
\end{equation}
in such a way that it commutes  with the fields $\partial_j$,
$j=1,\ldots, N-1$. This is verified provided
\begin{equation}\label{11}
\frac{\partial \ln X_i}{\partial
\gamma_j}=\frac{1}{\gamma_i-\gamma_j},\qquad i\neq j.
\end{equation}
By applying Tsarev's method (see \cite{martinez:7, martinez:8}) one finds

\begin{lemma}A general solution of \eqref{11} is
\begin{equation}\label{12}
D_x\gamma_i=\frac{a_i(\gamma_i)}{\prod\limits_{j\neq
i}(\gamma_i-\gamma_j)},\qquad i=1,\ldots,N,
\end{equation}
where $a_i(\xi)$ are $N$ arbitrary functions.
\end{lemma}

As a consequence we can
reformulate  \eqref{8} as a set of $N$ dynamical systems for
$\gamma_i$
\begin{gather}
\partial_n \gamma_i=\Omega_{ni}X_i, \qquad
n=1,\ldots,N-1,\nonumber\\
\partial_x \gamma_i=X_i.\label{13}
\end{gather}
Equivalently, we have
\begin{equation}\label{14}
\frac{\d \gamma_i}{a_i(\gamma_i)}=\frac{2}{\prod\limits_{j\neq
i}(\gamma_i-\gamma_j)}\left(\d x+\sum_{n=1}^{N-1} \Omega_{ni}\d
t_n\right),\qquad i=1,\ldots,N.
\end{equation}
One can explicitly invert \eqref{14} (see \cite{martinez:8}) and find
\begin{equation}\label{15}
(-1)^n\d t_n =\sum_{i=1}^N
\frac{\gamma_i^{N-n-1}}{a_i(\gamma_i)}\d \gamma_i,\qquad
n=0,\ldots,N-1,
\end{equation}
where we are denoting $t_0:=x$. Therefore, we conclude that the
general solution of~\eqref{8} is determined by the following
system of $N$ implicit relations
\begin{equation}\label{16}
\displaystyle (-1)^n
t_n=\sum_{i=1}^N\int^{\gamma_i}\frac{\gamma_i^{N-n-1}}{a_i(\gamma_i)}\d
\gamma_i,\quad n=0,\ldots,N-1.
\end{equation}

The  theorems which follow state some apparently new results on
the structure of the hierarchy~\eqref{1i}

\begin{theorem}
The differential forms
\begin{gather*}
\omega_1=g_1\d x+g_2\d t_1+\cdots+g_n\d
t_{n+1}+\cdots,\qquad (\mbox{normalized case}),\\
\omega_2=\frac{1}{g_0}\left(\d x-g_1\d t_1+\cdots-g_n\d
t_{n}+\cdots\right),\qquad (\mbox{Schwarzian case}),
\end{gather*}
are closed.
\end{theorem}
\begin{proof}
Let us consider the normalized case. Notice that
\[
\omega_1=\sum_{i\geq 0}{\rm Res}(\lambda^i G)\d t_i,
\]
where ${\rm Res}(A(\lambda))$ stands for the coefficient of
$\lambda^{-1}$ of Laurent series $A(\lambda)$ in $\lambda$.
Moreover, from~\eqref{1i}
\begin{equation}\label{17}
\frac{\partial }{\partial t_j}\,{\rm Res}(\lambda^i G)={\rm Res}\left(\lambda^i
\frac{\partial G}{\partial t_j}\right)={\rm Res}(\langle A_j,\lambda_i
G\rangle).
\end{equation}
On the other hand
\begin{gather*}
{\rm Res}(\langle A_j,\lambda_i G\rangle)={\rm Res}(\langle \lambda^j
G-(\lambda G)_-,\lambda_i G\rangle)=-{\rm Res}(\langle (\lambda^j G)_-
,(\lambda_i G)_+\rangle)\\
\phantom{{\rm Res}(\langle A_j,\lambda_i G\rangle)}{}
={\rm Res}(\langle A_i,(\lambda_j G)_-\rangle)={\rm Res}(\langle
A_i,\lambda_j G\rangle),
\end{gather*}
so that \eqref{17} implies
\[
\frac{\partial }{\partial t_j}\,{\rm Res}(\lambda^i G)=\frac{\partial
}{\partial t_i}\, {\rm Res}(\lambda^j G),
\]
which proves that $\omega_1$ is closed.

The proof for the Schwarzian case can be done similarly by writing
\[
\omega_2=\frac{1}{g_0}-\sum_{i\geq
1}\, {\rm Res}\left(\lambda^{i-1}\frac{G}{g_0}\right)\d t_i,
\]
and by taking into account that
\begin{equation}\label{18}
\frac{\partial }{\partial t_i}\frac{1}{g_0}=-\frac{\partial
}{\partial x}\frac{g_i}{g_0}.
\end{equation}
\end{proof}

As a consequence of this theorem there exist two functions
$q_i=q_i(x,{\boldsymbol{t}})$ verifying
\begin{equation}\label{19}
\omega_i=\d q_i,\qquad i=1,2,
\end{equation}
so that  we can rewrite the hierarchy \eqref{1i} as a system of
partial differential equations for~$q_i$.

\begin{theorem}
 Let $G^{(s)}=G^{(s)}(\lambda,x,{\boldsymbol{t}})$ be a solution of the
Schwarzian hierarchy~\eqref{1i}. If we define
$\hat{x}=q_2(x,{\boldsymbol{t}})$, where $q_2$ satisfies~\eqref{19}, then
\begin{equation}\label{tr}
G^{(n)}(\lambda,\hat{x},{\boldsymbol{t}}):=\frac{G^{(s)}(\lambda,x,{\boldsymbol{t}})}{g_0(x,{\boldsymbol{t}})},
\end{equation}
solves the normalized hierarchy~\eqref{1i}.
\end{theorem}

\begin{proof}
Observe that
\[
\frac{\partial \hat{x}}{\partial x}=\frac{1}{g_0},\qquad
\frac{\partial \hat{x}}{\partial t_i}=\frac{1}{g_i},\quad i\geq 1.
\]
Hence
\[
\frac{\partial x}{\partial \hat{x}}=g_0,\qquad \frac{\partial
x}{\partial t_i}=g_i,\quad i\geq 1.
\]
Therefore, by taking \eqref{18} into account, we have
\[
\frac{\partial G^{(n)}}{\partial t_i}=\left(\frac{\partial
}{\partial t_i}+g_i\frac{\partial }{\partial x}\right)
\frac{G^{(s)}}{g_0}=\frac{1}{g_0}\langle
A_i^{(n)}+g_i,G^{(n)}\rangle=\langle
A_i^{(s)},G^{(s)}\rangle_{\hat{x}},
\]
where
\[
A_i^{(n)}=\lambda^i+\cdots+\hat{g}_{i-1}\lambda+\hat{g}_i,\quad
\hat{g}_i:= \frac{g_i}{g_0},
\]
and $\langle
U,V\rangle_{\hat{x}}:=U\partial_{\hat{x}}V-V\partial_{\hat{x}}U$.
\end{proof}

\begin{remark}
 This theorem shows that a correspondence $G^{(s)}\rightarrow G^{(n)}$
between the Schwarzian and normalized hierarchies exists.
Reciprocally, a transformation $G^{(n)}\rightarrow G^{(s)}$ can
also be defined provided that the solution
$G^{(n)}=G^{(n)}(\lambda,x,{\boldsymbol{t}})$ does not vanish at $\lambda=0$.
To prove this statement let us define
\[
u:=\frac{1}{G^{(n)}|_{\lambda=0}},
\]
and the differential form
\[
\omega:=u\left(\d x+\sum_{i\geq 1}g_i\d t_i\right),
\]
where $g_n$ are the coefficients of the expansion of $G^{(n)}$ in
powers of $\lambda$. From \eqref{1i} it follows at once that
\[
\partial_i u=(ug_i)_x.
\]
Moreover, from the identities
\[
\partial_j A_i-\partial_i A_j=\langle A_j,A_i\rangle_,
\]
and taking into account that $g_i=A_i|_{\lambda=0}$, we have
\[
\partial_j g_i-\partial_i g_j=\langle g_j,g_i\rangle_,
\]
and therefore
\[
\partial_j(u g_i)=\partial_i(u g_j),
\]
which proves that $\omega$ is a closed form. It is now
straightforward to demonstrate that the function
\[
G^{(s)}(\lambda,\hat{x},{\boldsymbol{t}}):=u(x,{\boldsymbol{t}})G^{(n)}(\lambda,x,{\boldsymbol{t}}),
\]
where
\[
\hat{x}:=q(x,{\boldsymbol{t}}),\qquad \omega=\d q,
\]
verifies the Schwarzian hierarchy~\eqref{1i}.
\end{remark}

\section{Reductions and differential constraints}

In \cite{martinez:5} wide classes of reductions of the normalized hierarchy
\eqref{1i} were introduced by using the notion of compatible
differential constraints. They can be easily generalized to the
Schwarzian hierarchy as well. We will consider here reductions
defined by second-order differential constraints
\begin{equation}\label{22}
2G_{xx}G-G_x^2+4a(\lambda)-4U(\lambda,x,{\boldsymbol{t}})G^2=0,
\end{equation}
where $a(\lambda)$ is an arbitrary function and
\begin{equation}\label{23}
U(\lambda,x,{\boldsymbol{t}}):=\left(\frac{a(\lambda)}{G^2}\right)_+.
\end{equation}
Requiring compatibility between~\eqref{1i} and~\eqref{22} yields
\begin{equation}\label{14a}
\partial_n U=-\frac{1}{2}A_{n,xxx}+2UA_{n,x}+U_x A_n,\qquad n\geq
1.
\end{equation}

We notice that in terms of the generating function $H$ for the
conservation laws, equations~\eqref{22} and~\eqref{23} can be written
as
\begin{equation}\label{24r}
\{H,x\}+a(\lambda)H^2=U(\lambda,x,{\boldsymbol{t}}),\qquad
U=\left(a(\lambda)H^2\right)_+,
\end{equation}
where we are using the Schwarzian derivation operation
\[
\{H,x\}:=\frac{3}{4}\frac{H_x^2}{H^2}-\frac{1}{2}\frac{H_{xx}}{H}.
\]
The equations \eqref{22} determine integrable hierarchies
associated with Schr\"odinger spectral problems with
energy-dependent potentials. Thus, for $a(\lambda)=\lambda^n$ one
finds  the generalized KdV hierarchies for energy-dependent
potentials of the form
\begin{equation}\label{25}
\quad U(\lambda,x,{\boldsymbol{t}}):=\lambda^n+ \sum_{i=0}^{n-1}\lambda^i
u_i(x,{\boldsymbol{t}}),\qquad (\mbox{normalized case}),
\end{equation}
and
\begin{equation}\label{26}
\quad U(\lambda,x,{\boldsymbol{t}}):= \sum_{i=1}^{n}\lambda^i u_i(x,{\boldsymbol{t}}),\qquad
(\mbox{Schwarzian case}).
\end{equation}
For the normalized case $G=G^{(n)}$ the simplest choices $n=1$ and
$n=2$ lead to the KdV and the Zakharov--Shabat hierarchies,
respectively.  We observe that for the Schwarzian case the choice
$a(\lambda)=\lambda^n$ in \eqref{24r} leads to the constraint~\eqref{11i}.

\begin{theorem}
The differential constraint~\eqref{24r} is invariant under the
transformation
\linebreak  $H^{(s)}\rightarrow H^{(n)}$ determined by
equation~\eqref{tr}. That is to say, if $H^{(s)}$ satisfies
\[
\left\{H^{(s)},x\right\}+a(\lambda)H^{(s)2}=U^{(s)}(\lambda,x,{\boldsymbol{t}}),
\]
then
\[
\left\{H^{(n)},x\right\}+a(\lambda)H^{(n)2}=U^{(n)}(\lambda,x,{\boldsymbol{t}}).
\]
The corresponding transformation law for the potential function is
\begin{equation}\label{32}
U^{(n)}=g_0^2 U^{(s)}+\{g_0(\hat{x}),\hat{x}\}.
\end{equation}
\end{theorem}

\begin{proof}
If we define $H^{(n)}:=H^{(s)}/h_0$, then it follows at once that
\[
\left\{H^{(s)},x\right\}=h_0^2\left(
\left\{H^{(n)},\hat{x}\right\}+\frac{1}{4}\frac{h_{0,\hat{x}}^2}{h_0^2}-
\frac{1}{2}\frac{h_{0,\hat{x}\hat{x}}}{h_0}\right).
\]
On the other hand
\[
\left\{\frac{1}{h_0},\hat{x}\right\}
=\frac{1}{2}\frac{h_{0,\hat{x}\hat{x}}}{h_0}-\frac{1}{4}\frac{h_{0,\hat{x}}^2}{h_0^2}.
\]
Thus, it is straightforward to get
\begin{equation}\label{33}
\left\{H^{(n)},\hat{x}\right\}+a(\lambda)H^{(n)2}=
U^{(n)},\quad U^{(n)}:=\frac{1}{h_0^2}
U^{(s)}+\left\{\frac{1}{h_0},\hat{x}\right\}.
\end{equation}
\end{proof}

A complete and convenient description of the general
class of the \emph{energy-dependent} hierarchies can be formulated
by taking advantage of the identities~\eqref{24r}. The following
examples illustrate the fact that, in addition to the standard
hierarchies of Zakharov--Shabat and KdV types, many other
integrable models may be analyzed within the theory of the second
order differential constraints  of the Schwarzian hierarchy~\eqref{1i}.

\begin{example}  For $a(\lambda)=\lambda$ and
   $G=G^{(s)}$, the constraint \eqref{24r} reads
\[
 \{H,x\}+\lambda H^2=\lambda u_1,
\]
and by identifying the coefficient of $1/\lambda$ we get
\[
\{h_0,x\}+2h_0h_1=0.
\]
Hence, by taking into account that in terms of the function
$q:=q_2(x,{\boldsymbol{t}})$ of \eqref{19}
\begin{equation}\label{27}
h_0=\frac{1}{g_0}=q_x,\qquad h_1=-\frac{g_1}{g_0^2}=q_xq_t,\qquad
(t:=t_1),
\end{equation}
we deduce at once the Schwarzian form of the KdV equation
\begin{equation}\label{28}
2q_tq_x=\{q_x,x\}\Leftrightarrow-8q_{t}+2\frac{q_{xxx}}{q_x^3}-3\frac{q_{xx}^2}{q_x^4}=0,
\end{equation}
which, written in terms of $g_0=1/h_0$, reduces to the  Harry--Dym
equation
\begin{equation}\label{29}
g_{0,t}=\frac{1}{4}g_0^3g_{0,xxx}.
\end{equation}
\end{example}

\begin{example}
 For $a(\lambda)=\lambda^2$ and
$G=G^{(s)}$, the identity \eqref{24r} reads
\[
 \{H,x\}+\lambda^2 H^2=\lambda^2u_2+\lambda u_1,
\]
and by identifying the constant term in the expansion in powers
of $\lambda$ we get
\[
\{h_0,x\}+h_1^2+2h_0h_2=0.
\]
Now, from \eqref{4i} we have
\[
h_{0,t}=\left(\frac{h_1}{h_0}\right),\qquad
h_{1,t}=\left(\frac{h_2}{h_0}\right),\qquad (t:=t_1).
\]
 Hence, by
taking into account \eqref{27} we find
\begin{equation}\label{30}
(q_xq_t)_t=\left(\frac{1}{4}\frac{q_{xxx}}{q_x^3}-\frac{3}{8}\frac{q_{xx}^2}{q_x^4}-
\frac{1}{2}q_t^2\right)_x.
\end{equation}
This nonlinear equation admits a Lagrangian given by
\begin{equation}\label{31}
\mathfrak{L}=\iint
\left(q_t^2q_x+\frac{1}{4}\frac{q_{xx}^2}{q_x^3}\right)\d x\d t,
\end{equation}
or, equivalently, under the point transformation $q(u(y,t),t)=y$
\begin{equation}\label{31a}
\mathfrak{L}=\iint\frac{u_t^2+u_{yy}^2}{u_y^2}\d y\d t.
\end{equation}
If we denote now $y\rightarrow x$, $u\rightarrow q$, the following
equivalent Lagrangian formulation arises
\begin{gather*}
 \mathfrak{L}=\iint\left(\frac{q_t^2+q_{xx}^2}{q_x^2}\right)\d x\d t
=\iint\left(\frac{(q_t+q_{xx})^2}{q_x^2}-
2\frac{q_tq_{xx}}{q_x^2}\right)\d x\d
t\\
\phantom{\mathfrak{L}}{}=\iint\frac{(q_t+q_{xx})^2}{q_x^2}\d x\d t.
\end{gather*}
Thus, the model admits a Hamiltonian density
\[
 H=q_tL_{q_t}-L=p_xq_x+\frac14p^2q_x^2,\qquad p:= L_{q_t},\qquad
 L:=\frac{(q_t+q_{xx})^2}{q_x^2},
\]
which leads to the Hamiltonian equations
\[
q_t+q_{xx}+2pq_x^2=0,\qquad -p_t+p_{xx}=(2p^2q_x)_x.
\]
It should be noticed that the last system is one of ``canonical''
forms of the isotropic Landau--Lifshitz model (see~\cite{martinez:9}).
\end{example}

In \cite{martinez:5} first-order differential constraints defining reductions of
\eqref{1i} were also introduced. They adopt the form
\begin{equation}\label{34}
G_x-a(\lambda)-U(\lambda,x,{\boldsymbol{t}})G=0,
\end{equation}
where $a(\lambda)$ is an arbitrary function and
\begin{equation}\label{35}
U(\lambda,x,{\boldsymbol{t}}):=-\left(\frac{a(\lambda)}{G}\right)_+.
\end{equation}
Requiring compatibility between \eqref{1i} and \eqref{34} gives
\begin{equation}\label{10a}
 \partial_n U=- A_{n,xx}+(UA_n)_x,\qquad n\geq 1,
\end{equation}
or, equivalently,
\begin{equation}\label{10b}
\partial_n U=-\partial_x(U(\lambda^n G)_-)_+,\qquad n\geq 1,.
\end{equation}
In terms of the function $H$ equations~\eqref{34} and~\eqref{35} read
\begin{equation}\label{34a}
H_x+a(\lambda)H^2+U(\lambda,x,{\boldsymbol{t}})H=0,\qquad
U=-\Big(a(\lambda)H\Big)_+.
\end{equation}
We notice that for the Schwarzian case if we set
$a(\lambda)=-\lambda^m$ in~\eqref{34} the constraint~\eqref{10i}
follows.

First-order differential constraints determine in turn further
reductions of the second-order differential constraints. Indeed,
it follows easily that

\begin{theorem}
Given a solution $G$ of \eqref{34} then it satisfies
\[
2G_{xx}G-G_x^2+a(\lambda)^2-4\widehat{U}(\lambda,x,{\boldsymbol{t}})G^2=0,
\]
where $U\rightarrow \widehat{U}$ is implemented by the Miura
transformation
\[
\widehat{U}:=\frac{1}{2}U_x+\frac{1}{4}U^2.
\]
\end{theorem}

Among the reductions defined by first-order differential
constraints one finds the Bur\-gers hierarchy as well as a class of
hierarchies of \emph{energy-dependent} type.

\begin{example}
If we impose on $G=G^{(n)}$ the
constraint
\[
G_x=(\lambda+u)G-\lambda,\qquad u=h_1=-g_1,
\]
we get the Burgers hierarchy
\[
\partial_n u=\partial_x\left((\partial_x+u)^n u\right).
\]
In particular  the flow corresponding to $t:=t_1$ is the Burgers
equation
\[
u_t=u_{xx}+2uu_x.
\]
\end{example}

\begin{example}
For a constraint on $G=G^{(n)}$
of the form
\begin{equation}\label{35a}
G_x=(\lambda^2+\lambda u_1+u_0)G-\lambda^2,\qquad
u_1=h_1=-g_1,\qquad u_0=h_2=g_1^2-g_2,
\end{equation}
 we get the following hierarchy for the \emph{energy-dependent}
potential $U:=\lambda^2+\lambda u_1+u_0$
\begin{equation}\label{36}
\partial_n\left (\begin{array}{l}
u_1\\u_0
\end{array}\right )=
\left (\begin{array}{cc}
-\partial_x & 0\\
0 & -\partial_x(\partial_x-u_0)\end{array}\right )R^{n+1}\left
(\begin{array}{l} 1\\0
\end{array}\right) ,
\end{equation}
where
\[
 R:=\left (\begin{array}{cc}
-u_1&\partial_x +u_0\\
1 & 0\end{array}\right ).
\]
The flow corresponding to $t:=t_1$ reads
\begin{equation}\label{37}
u_{0,t}=u_{1,xx}-(u_0u_1)_x,\qquad u_{1,t}=-\left(u_1^2+u_0\right)_x,
\end{equation}
which in terms of the function $q:=q_1(x,{\boldsymbol{t}})$ of~\eqref{19}
reduces to
\[
q_{tt}-q_{xxx}-3q_xq_{xt}-q_tq_{xx}+(q_x)^3_x=0.
\]
\end{example}

 It is  possible to characterize multidimensional integrable
models by imposing compatible differential constraints
to~\eqref{1i}. To illustrate this feature let us consider the
constraint~\eqref{35a} for $G=G^{(n)}$ and denote $y:=t_1$, $t:=t_2$.
Thus we have the system of equations
\begin{subequations} \label{eqA}
\begin{gather}
  G_x=\left(\lambda^2-\lambda g_1+g_1^2-g_2\right)G-\lambda^2,\label{eqAa}\\
  G_y=\langle \lambda+g_1,G\rangle ,  \label{eqAb}\\
  G_t=\langle \lambda^2+\lambda g_1+g_2 ,G\rangle  .
  \label{eqAc}
\end{gather}
\end{subequations}
Let us introduce the function
\[
  F:=u\,G=f_0+\frac{f_1}{\lambda}+\frac{f_2}{\lambda^2}+\cdots,\qquad
  u:=\frac{1}{G|_{\lambda=0}}.
\]
From \eqref{eqA} we get
\begin{gather*}
  u_x=\left(g_2-g_1^2\right)u,\\
  u_y=\left(g_{1,x}+g_1g_2-g_1^3\right)u,  \\
  u_t=\left(g_{2,x}+g_2^2-g_2g_1^2\right)u ,
\end{gather*}
so that the equations \eqref{eqA} written in terms of $F$ read
\begin{subequations} \label{eqC}
\begin{gather}
  F_x=\left(\lambda^2-\lambda g_1\right)F-\lambda^2f_0,\label{eqCa}\\
  F_y=\left(\lambda^3-\lambda g_2\right)F-\lambda^3 f_0-\lambda f_1 ,  \label{eqCb}\\
  F_t=\left(\lambda^4-\lambda g_3\right)F-\lambda^4 f_0-\lambda^3 f_1-\lambda^2 f_2  .
  \label{eqCc}
\end{gather}
\end{subequations}
Now it is easy to deduce from \eqref{eqC} that
\begin{equation}\label{37a}
F_t=F_{xx}+2g_1F_y-g_1^2 F_x,
\end{equation}
and, as a consequence, one finds that $v:=g_1$ and $w=\ln u$
satisfy the nonlinear system of equations
\begin{gather}
v_t=v_{xx}+2vv_y-v^2v_x+2v_xw_x,\nonumber\\
w_t=w_{xx}+2vw_y-v^2w_x+w_x^2.\label{37b}
\end{gather}

We observe that equations~\eqref{eqC} mean that equation~\eqref{37a} and,
consequently, equation~\eqref{37b} admit separation of variables. If we
expand $F$ in powers of $\lambda$ then equations~\eqref{eqC} become
\begin{subequations} \label{eqD}
\begin{gather}
  f_{n,x}=f_{n+2}-g_1f_{n+1},\qquad f_{-1}=0,\qquad f_0=u,\label{eqDa}\\
  f_{n,y}=f_{n+3}-g_2f_{n+1},\qquad f_2=ug_2,  \label{eqDb}\\
  f_{n,t}=f_{n+4}-g_3f_{n+1} ,\qquad f_3=ug_3 .
  \label{eqDc}
\end{gather}
\end{subequations}
By using \eqref{eqDa} all the coefficients $f_n$ can be expressed
as differential polynomials in $u$ and $v=g_1$:
\[
f_0=u,\quad f_1=uv,\qquad f_2=u_x+uv,\qquad
f_3=(uv)_x+vu_x+uv^3,\qquad \ldots
\]

   Furthermore, the system \eqref{eqD} implies that polynomial reductions of
   the form
\[
 f_n=0,\qquad \forall \; n\geq N,
\]
are Liouville integrable. Indeed, by noticing that
$f_{N-1}=\mbox{const}$, we require $N-1$ commuting vector fields
to integrate such a reduction. For example, for $N=4$ the
dynamical variables are $f_0$, $f_1$, $f_2$ and~\eqref{eqD} can be
written as
\begin{gather*}
\d f_0=\left(f_2-\frac{f_1^2}{f_0}\right)\d
x+\left(\alpha-\frac{f_1f_2}{f_0}\right)\d y-\alpha\frac{f_1}{f_0}\d
t,\\
\d f_1=\left(\alpha-\frac{f_1f_2}{f_0}\right)\d x-\frac{f_2^2}{f_0}\d
y-\alpha\frac{f_2}{f_0}\d t,\\
\d f_2=-\alpha\frac{f_1}{f_0}\d x-\alpha\frac{f_2}{f_0}\d
y-\frac{\alpha^2}{f_0}\d t,
\end{gather*}
where $\alpha_3:=f_3=\mbox{const}$. It is now straightforward to
eliminate $\d x$, $\d y$, $\d t$ and find
\begin{gather*}
\d x=\d\left( \frac{f_1}{\alpha}-\frac{f_2}{2\alpha^2}\right),\\
\d y=\d
\left(\frac{f_0}{\alpha}-\frac{f_1f_2}{\alpha^2}+\frac{f_2^3}{3\alpha^3}\right),\\
\d t=\d \left(-\frac{f_0f_2}{\alpha^2}-\frac{f_1^2}{2\alpha^2}+
\frac{f_1f_2^2}{\alpha^3}-\frac{f_2^4}{4\alpha^4}\right),
\end{gather*}
In this way the solution is determined by the following system
\begin{gather*}
f_0=\alpha y+xf_2-\frac{f_2^3}{2\alpha},\qquad f_1=\alpha x+\frac{f_2^2}{2\alpha}, \\
-\frac{7}{8}f_2^4+\frac{1}{2}\alpha^2 xf_2^2+\alpha^3
f_2+\alpha^4\left(\frac{1}{2}x^2+t\right)=0.
\end{gather*}
This example shows the complicated algebraic singularities
exhibited by the solutions of~\eqref{37b}.

\subsection*{Acknowledgements}

A~B~Shabat wish to thank  A~Ibort for his support during his stay
as a visiting professor of the Carlos III University of Madrid. He
also acknowledges the group  of the Department of Theoretical
Physics II of the Complutense University for their warm
hospitality.  L~Martinez Alonso was partially supported by the
DGCYT project BFM2002-01607. A~Shabat  was partially supported by
the RFFI project  01-01-00874 and Sc. Schools RF grant
06-15-96093. Both authors are grateful to E~Ferapontov for several
useful and stimulating discussions.

\label{martinez-lastpage}
\end{document}